\documentclass[aip,jap,reprint,twocolumn]{revtex4-2}

\usepackage{etoolbox,graphicx,amsmath,amssymb}
\usepackage[utf8]{inputenc}
\usepackage[T1]{fontenc}
\usepackage[english]{babel}

\makeatletter
\def\@email#1#2{
 \endgroup
 \patchcmd{\titleblock@produce}
 {\frontmatter@RRAPformat}
 {\frontmatter@RRAPformat{\produce@RRAP{*#1\href{mailto:#2}{#2}}}\frontmatter@RRAPformat}{}{}}
\makeatother

\begin{document}

\title{Graphene-based enhancement of near-field radiative-heat-transfer rectification}

\author{Simon Landrieux}
\affiliation{Université Paris-Saclay, Institut d'Optique Graduate School, CNRS, Laboratoire Charles Fabry, 91127, Palaiseau, France}
\author{Philippe Ben-Abdallah}
\affiliation{Université Paris-Saclay, Institut d'Optique Graduate School, CNRS, Laboratoire Charles Fabry, 91127, Palaiseau, France}	
\author{Riccardo Messina}
\email{riccardo.messina@institutoptique.fr}
\affiliation{Université Paris-Saclay, Institut d'Optique Graduate School, CNRS, Laboratoire Charles Fabry, 91127, Palaiseau, France}

\date{\today}

\begin{abstract}
We present a thermal device based on the near-field interaction between two substrates made of a polar and a metal-insulator-transition material. As a result of the temperature dependence of the optical properties, this device acts as a thermal rectifier, implying a strong asymmetry in the heat flux when reversing the two temperatures. By covering both substrates with a graphene sheet we show a significant enhancement of rectification coefficient. By investigating the flux spectral properties along with its distance dependence, we prove that this enhancement is associated to a change in the power-law dependence of heat flux with respect to the separation distance in the electrostatic regime due to the presence of graphene sheets. Our results highlight the promising role of graphene-based hybrid structures in the domain of nanoscale thermal management.
\end{abstract}

\maketitle

Two bodies at different temperatures exchange energy under the form of thermal radiation even when separated by vacuum. This radiative heat transfer (RHT) is well described at large separation distances by Stefan-Boltzmann's law, setting an upper bound to this energy exchange, reached only by two blackbodies. During the second half of the 20\textsuperscript{th} century, the works of Rytov~\cite{Rytov89}, Polder and van Hove~\cite{Polder71} and the development of fluctuational electrodynamics showed that this limit does not apply in the near field, i.e. for separation distances smaller than the thermal wavelength $\lambda_\text{th}=\hbar c/k_B T$ (some microns at ambient temperature). In this regime, the flux can overcome even by several orders of magnitude the blackbody limit, especially in the presence of resonant surface modes of the electromagnetic field, such as surface phonon-polaritons for polar materials~\cite{Joulain05}. Since then, these theoretical predictions have been confirmed by a large number of experiments, stimulating in turn further theoretical investigations (see Refs.~\onlinecite{Cuevas18,Song15,Biehs21} and references therein).

These developments have paved the way to the exploration of the role played by near-field RHT in several applications, including heat-assisted data recording~\cite{Challener09,Stipe10}, infrared spectroscopy~\cite{DeWilde06,Jones12}, energy-conversion techniques~\cite{DiMatteo01,Narayanaswamy03,Basu07,Fiorino18} and thermotronics~\cite{Abdallah13a,Abdallah15}, i.e. the design of thermal equivalent of circuit elements. A significant attention has been devoted, both theoretically~\cite{Otey10,Basu11,Zwol11,Iizuka12,Wang13,Zhu13,Yang13,Huang13,Joulain15,Gu15,Yang15,Ghanekar16,Tang17,Ordonez17,Zheng17,Ghanekar18,Shen18,Xu18b,Xu18a,Wen19,Ott19,Toyin21,Li21,Moncada21,Feng21,Liu21,Chen21,Latella21} and experimentally~\cite{Zwol12,Ito17,Elzouka17,Fiorino18}, to the rectification of RHT in the near field, consisting in a configuration implying a strong asymmetry of heat flux when exchanging the temperatures of the two bodies. In a two-terminal system, this effect is due to a temperature dependence of the optical properties of the materials, and has been extensively studied both for planar and non-planar systems~\cite{Zhu13,Wen19,Toyin21}, from cryogenic to ambient and high temperatures. This dependence is not required in systems made of more than two objects~\cite{Latella21}, in which rectification can result from purely many-body effects. The sensitivity of the rectification coefficient to a variety of aspects, such as film thickness~\cite{Yang15,Tang17,Li21}, doping~\cite{Basu11}, dielectric coating~\cite{Iizuka12}, gratings~\cite{Ghanekar16,Ghanekar18,Shen18,Liu21,Chen21} has been explored. Concerning the choice of materials, while phase-transition materials~\cite{Zwol11,Zwol12,Huang13,Yang13,Gu15,Ghanekar16,Ito17,Zheng17,Ghanekar18,Fiorino18,Liu21,Li21,Chen21} such as vanadium dioxide (VO$_2$) have gathered a remarkable attention, also superconductors~\cite{Ordonez17,Moncada21} and graphene-based systems~\cite{Zheng17,Xu18a,Xu18b} have been investigated. While it is often difficult to compare different near-field rectification schemes due to strongly different geometrical and physical features (see Ref.~\onlinecite{Wen19} for a summary of recent results), the most recent works~\cite{Wen19,Chen21,Feng21,Li21,Liu21,Moncada21} propose scenarios with an efficiency, defined as $\eta = \frac {\Phi_\text{max} -\Phi_\text{min}} {\Phi_\text{max}}$ ($\Phi_\text{max}$ and $\Phi_\text{min}$ being the maximum and minimum fluxes, respectively), going beyond 90\%, and limited mainly by the intrinsic losses of the materials and the mismatch between their surface resonances.

In this work we address the role played by graphene in modulating and enhancing the efficiency of a near-field radiative thermal rectifier based on two parallel substrates made of VO$_2$ and silica (SiO$_2$). Graphene has recently proved to be promising for near-field RHT, since it supports a delocalized surface plasmon~\cite{Ilic12a} which can be used to improve the coupling between two different materials and whose properties can be modulated by acting on the graphene chemical potential. Based on this behavior and on the coupling between graphene plasmons and surface modes of an underlying substrate~\cite{Messina13a}, the possibility of tuning the heat transfer and increasing the performances of energy-conversion devices has been investigated~\cite{Svetovoy12,Ilic12b,Lim13,Messina13b,Svetovoy14,Messina17,Papadakis19}. In this Letter, we propose a system in which a graphene sheet is deposed both on the VO$_2$ and the SiO$_2$ films. By optimizing for each distance the chemical potentials of the two sheets, we show a modulation of the efficiency of the rectification mechanism and an increase by a maximum of 14\% for a vacuum gap of 100\,nm. By studying the dependence on the separation distance and the graphene chemical potential along with the flux spectral features, we highlight the crucial role played by graphene plasmons in the increased coupling between the two films. The main difference between our results and previous works~\cite{Zheng17,Xu18a,Xu18b} focusing on graphene is both the simultaneous use of two graphene sheets and the optimization of the chemical potentials for any separation distance, allowing us to maximize the coupling between dissimilar materials exchanging heat in the near field. Moreover, we highlight a non-monotonic behavior of the efficiency enchancement ascribed to a transition in the power-law behavior of RHT. These results pave the way to the active manipulation of heat flux at the nanoscale.

\begin{figure}
	\centering
	\includegraphics[width=0.48\textwidth]{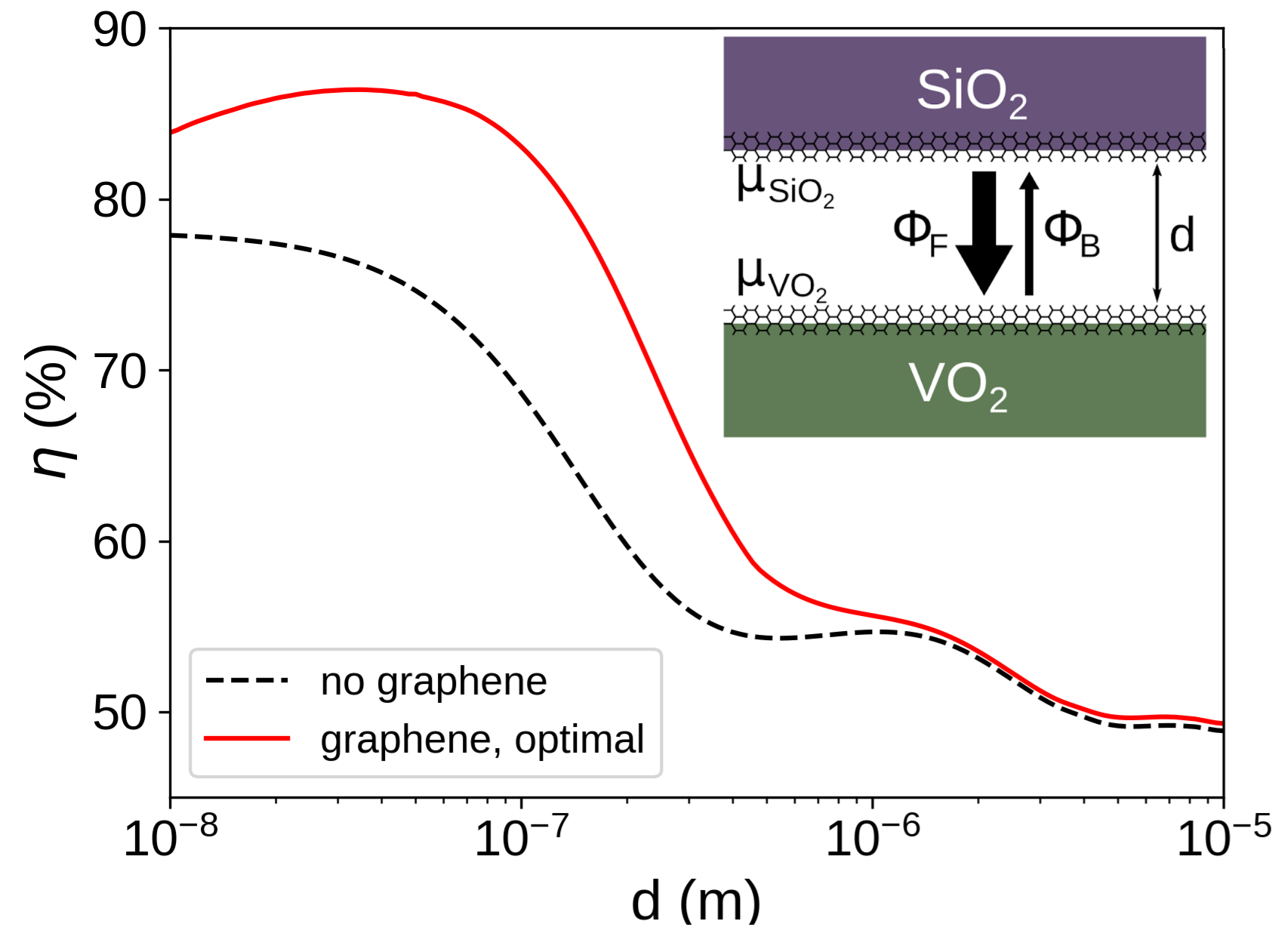}
	\caption{Rectification efficiency $\eta$ for the thermal diode without graphene (dashed black line) or with the graphene sheets tuned to the chemical potentials that maximize $\eta$ (red solid line) as a function of the separation distance $d$. The calculations were done for $T_{\text{high/low}}=T_c \pm 100\,$K.}\label{fig:eta}
\end{figure}

The radiative thermal diode (RTD) we consider, depicted in the inset of Fig.~\ref{fig:eta}, consists of two semi-infinite substrates separated by a vacuum gap of width $d$, one made of vanadium dioxyde (VO$_2$) and the other made of fused silica (SiO$_2$), both covered with a monolayer graphene sheet at a chemical potential $\mu_{\text{VO}_2}$ and $\mu_{\text{SiO}_2}$, respectively. For a given temperature gradient $\Delta T = T_{\text{high}}-T_{\text{low}}$, $\Phi_F$ is the heat flux received by VO$_2$ at $T_{\text{low}}$ from SiO$_2$ at $T_{\text{high}}$, whereas $\Phi_B$ is the one received by SiO$_2$ at $T_{\text{low}}$ from VO$_2$ at $T_{\text{high}}$. The rectification coefficient $\eta = \frac {\Phi_F -\Phi_B} {\Phi_F}$ is used to assess the efficiency of the system as a thermal diode.

It is consequently necessary to have $\Phi_F \gg \Phi_B$ in order to have $\eta \simeq 1$. This is achieved (with or without graphene) by exploiting the first-order Mott transition of VO$_2$~\cite{Qazilbash07} from a low-temperature insulating phase to a high-temperature metallic one close to room temperature ($T_c = 340\,$K). Indeed, if $T_c \in [T_{\text{low}},T_{\text{high}}]$, VO$_2$ is dielectric at $T_{\text{low}}$ and metallic at $T_{\text{high}}$, considerably reducing the transmission coefficient in the latter case due to the absence of a resonant mode in the infrared range. This results in a $\Phi_B$ significantly lower than $\Phi_F$, hence in a high $\eta$. However, the use of a second material different from VO$_2$ to break the symmetry of the system implies a mismatch between the frequencies of the surface modes of the two materials. Due to the crucial role they play in the RHT at the nano-scale, this has a negative effect on $\Phi_F$ and consequently on $\eta$.

The way to enhance the rectification coefficient $\eta$ of the RTD we present consists in increasing $\Phi_F$ by adjusting the chemical potentials $\mu_{\text{VO}_2}$ and $\mu_{\text{SiO}_2}$ of the two graphene sheets to tune the surface modes of the two sides of the system. An upper limit of $\mu = 0.5\,$eV has been set for the calculations to be coherent with experimentally attainable values. We compute the heat fluxes $\Phi_{\text{F},\text{B}}$ between the two bodies by means of the following Landauer-like decomposition~\cite{Abdallah10,Biehs10}:
\begin{equation}\Phi = \int_0^\infty\frac {d \omega} {2 \pi} \Delta \Theta (\omega)\sum_p\int\frac {d^2 \mathbf{k}} {(2 \pi)^2} \mathcal T_p(\omega, \mathbf{k}),\label{eq:RHT}\end{equation}
where $\Delta \Theta (\omega) = \Theta(\omega,T_{\text{high}}) - \Theta(\omega,T_{\text{low}})$ is the difference in mean energy of a harmonic oscillator $\Theta(\omega,T)=\hbar\omega/[\exp(\hbar\omega/k_BT)-1]$ at the two temperatures, $\mathbf{k}$ is the projection of the wave vector on the surface of the semi-spaces and $\mathcal T_p$ is the Landauer transmission coefficient for the polarization $p$ [transverse electric (TE) or magnetic (TM)], taking values between 0 and 1, giving the proportion of the energy discrepancy $\Delta\Theta$ between the modes of the two sides that is actually transferred from the hot to the cold side. For two parallel semi-infinite planar bodies separated by a vacuum gap of width $d$, it is given by
\begin{equation}
\mathcal T_p(\omega,\mathbf{k}) = 
\begin{cases}
\displaystyle \frac {(1-|r_{1,p}|^2)(1-|r_{2,p}|^2)} {|1-r_{1,p}r_{2,p}e^{2 i k_z d}|^2} ,& k< \omega /c \\
\displaystyle \frac {4 \ \text {Im}(r_{1,p}) \text {Im}(r_{2,p}) e^{-2\text {Im} (k_z)d}}{|1 - r_{1,p}r_{2,p} e^{-2 \text {Im} (k_z)d}|^2} ,&  k> \omega /c
\end{cases}
\label{eq:T}\end{equation}
where $k_z = \sqrt {(\omega /c)^2- \mathbf{k}^2}$ is the component of the wave vector normal to the surfaces, $r_{i,p}$ is the reflection coefficient for medium $m$ at polarization $p$, which implicitly depends on $\omega$ and $k$. In the case of a graphene-covered surface, the reflection coefficients read~\cite{Messina13b}:
\begin{equation}
\begin{split}
r_{i,\text{TE}} &= \displaystyle \frac {k_z - k_{z,m} -\mu_0 \omega \sigma(\omega)} {k_z + k_{z,m} + \mu_0 \omega \sigma(\omega)}, \\
r_{i,\text{TM}} &= \displaystyle \frac {\epsilon _m (\omega) k_z - k_{z,m} + \frac {1} {\epsilon _0 \omega} \sigma(\omega)k_z k_{z,m} } {\epsilon _m (\omega) k_z + k_{z,m} + \frac {1} {\epsilon _0 \omega} \sigma(\omega)k_z k_{z,m}},
\end{split}
\label{eq:r}\end{equation}
where $\epsilon _m (\omega)$ is the dielectric permittivity of medium $m$, $k_{z, m} = \sqrt {(\omega /c)^2\epsilon _m (\omega)-\mathbf{k}^2 }$ is the normal component of the wave vector inside medium $m$ and $\sigma(\omega)$ is the conductivity of graphene. The latter is typically expressed as the sum of an intraband (Drude) contribution $\sigma_D (\omega)$ and an interband contribution $\sigma_I (\omega)$, which are given by the following expressions~\cite{Falkovsky08}:
\begin{equation}
\begin{split}
& \sigma_D (\omega) = \frac {i} {\omega + i / \tau} \frac {2e^2k_BT} {\pi \hbar ^2} \log \Big[2\cosh \Big (\frac {\mu} {2 k_B T}\Big) \Big] \text{,} \\
& \sigma _I (\omega) = \frac {e^2} {4\hbar} \Big[G \Big(\frac {\hbar \omega} 2\Big) + i \frac {4 \hbar \omega} \pi \int _0 ^\infty \frac {G(\xi)-G(\frac {\hbar \omega} 2)} {(\hbar \omega)^2 - 4 \xi ^2} d \xi \Big] \text{,}
\end{split}
\label{eq:sigma}\end{equation}
where $\mu$ is the chemical potential of the graphene, $\tau$ is the relaxation time, for which we take a value of $10 ^{-13}\,$s (see Refs.~\onlinecite{Ilic12b,Messina13b}), and $G(x) = {\sinh \frac x {k_B T}} /(\cosh \frac x {k_B T} + \cosh \frac \mu {k_B T})$. As clear from Eq.~\eqref{eq:sigma}, the chemical potential $\mu$ has an influence on the conductivity of graphene, resulting in the possibility to tune, by controlling (chemically or electrically) $\mu$, the Landauer coefficient $\mathcal T _p$ and thus the RHT between the two bodies.

In order to numerically highlight the enhancement of rectification efficiency offered by the presence of graphene, we have calculated the forward and backward fluxes for $T_{\text{high/low}}=T_c \pm 100\,$K in two configurations, namely in the absence and in the presence of graphene. In the latter scenario, for each distance $d$ between 10\,nm and 10\,$\mu$m, we have found the values of $\mu_{\text{VO}_2}$ and $\mu_{\text{SiO}_2}$ maximizing the efficiency $\eta$. Concerning the permittivities, for SiO$_2$ we calculate it by interpolating experimental data~\cite{Palik98}, while for VO$_2$, the dielectric phase is approximated by an isotropic medium of dielectric constant $\epsilon _{\text{VO}_2, \text{d}}(\omega)$ computed by a classical oscillator model and the metallic phase is described by a Drude-Lorentz model~\cite{Barker66}. The enhancement of the efficiency $\eta$ resulting from the presence of graphene is illustrated in Fig.~\ref{fig:eta}. We clearly observe, for values of $d$ lower than $4\,\mu$m, an increase of the efficiency, becoming negligible for larger distances. More specifically, the graphene-enhanced RTD has a maximal rectification coefficient of $\eta = 85 \%$ for $d \simeq 100\  \text{nm}$, whereas the one without graphene has $\eta = 71 \%$ for the same $d$. Moreover, the adjunction of graphene enables to attain values of $\eta$ $6 \%$ higher than the absolute maximum ($\eta\simeq78\%$) of the efficiency without graphene, values which would be hard to achieve due to the small gap widths $d$. On the other hand, at closer separation distance, we observe that the increase of rectification coefficient becomes less pronounced. As discussed below, this result can be ascribed to the behavior of graphene in the electrostatic regime.

It is well known~\cite{Messina13b} that the plasmons of graphene play a relevant role for relatively high values of the parallel wave-vector $k$~\cite{Messina13b}. This explains the impossibility to enhance the efficiency at distances above 1\,$\mu$m. On the contrary, in order to get further insight into the mechanism behind the efficiency amplification taking place at smaller separation distances, we focus on the configuration $d = 100\,$nm, both without graphene and with graphene at the chemical potentials $\mu_{\text{VO}_2} = 0.4\,$eV and $\mu_{\text{SiO}_2} = 0.2\,$eV for VO$_2$ and SiO$_2$, respectively, corresponding to the configuration maximizing $\eta$ at $d = 100\,$nm.

\begin{figure}
	\centering
	\includegraphics[width=0.48\textwidth]{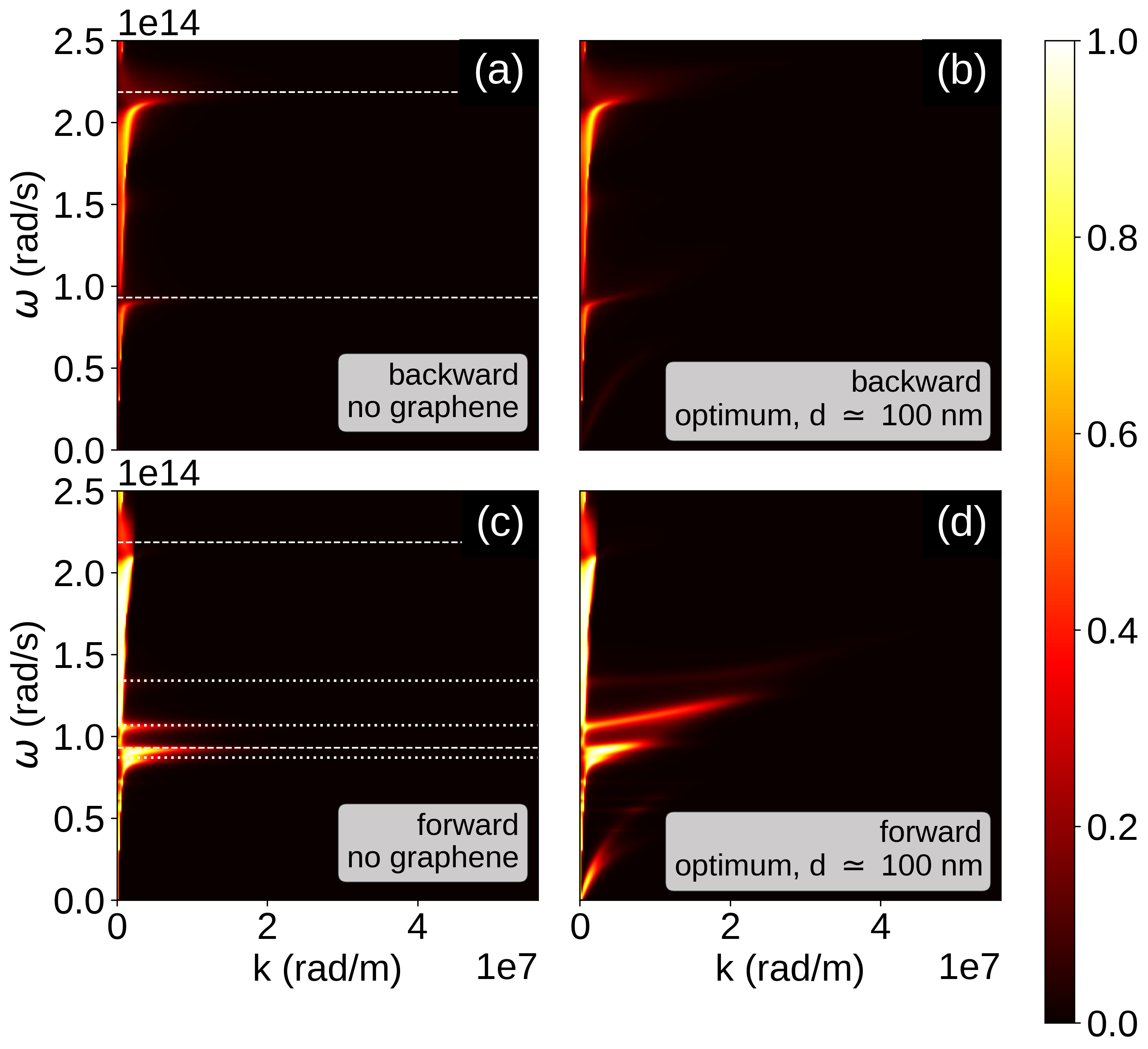}
	\caption{Transmission coefficient $\mathcal{T}_{\text{TM}} (\omega, \mathbf{k})$ in the $(k, \omega)$ plane for the backward temperature gradient [(a), (b)] and for the forward temperature gradient [(c), (d)]. The left column [(a), (c)] correponds to the RTD without graphene, while the right column [(b), (d)] correponds to the RTD with graphene at the optimal chemical potentials for $d \simeq 100\,$nm. The horizontal dashed [(a)] and dotted (a) and (c)] lines correspond to the surface resonances of SiO$_2$ and VO$_2$ (dielectric phase), respectively.}\label{fig:tplane}
\end{figure}

The participation of the surface resonant modes to the RHT in the two configurations are represented in Fig.~\ref{fig:tplane}, where the transmission coefficient in TM polarization $\mathcal{T}_{\text{TM}} (\omega, \mathbf{k})$, the one contributing the most to the RHT, are represented in the $(k, \omega)$ plane. In the first row (corresponding to the backward temperature gradient), the main resonant modes observable are the ones of SiO$_2$ at frequencies $\omega_{\text{SiO}_2,1} = 0.9 \times 10^{14}\,$rad/s and $\omega_{\text{SiO}_2,2} = 2.3 \times 10^{14}\,$rad/s, both without (a) and with graphene (b). In the optimal configuration with graphene (b), a third branch of low intensity for low $\omega$ is barely visible, and corresponds to the surface plasmon of graphene alone. We conclude that the adjunction of graphene has little effect on the transmission coefficients for the backward temperature gradient due to the absence of resonant modes in the Planck window for VO$_2$ in its metallic phase. In the second row (forward temperature gradient), a multitude of resonant modes are observable for the RTD without graphene [panel (c)] in the $[0.8,1.1] \times 10^{14}\,$rad/s range, which can be linked to resonant surface modes of VO$_2$ in its dielectric phase. For the optimal configuration with graphene [panel (d)], while the signature of the modes of graphene alone can still be seen for low $\omega$, the most noticeable effect of graphene is its coupling with the modes of the two substrates, which results in the broadening of their frequency domain and in an increased participation of modes with higher $k$.

\begin{figure}
	\centering
	\includegraphics[width=0.48\textwidth]{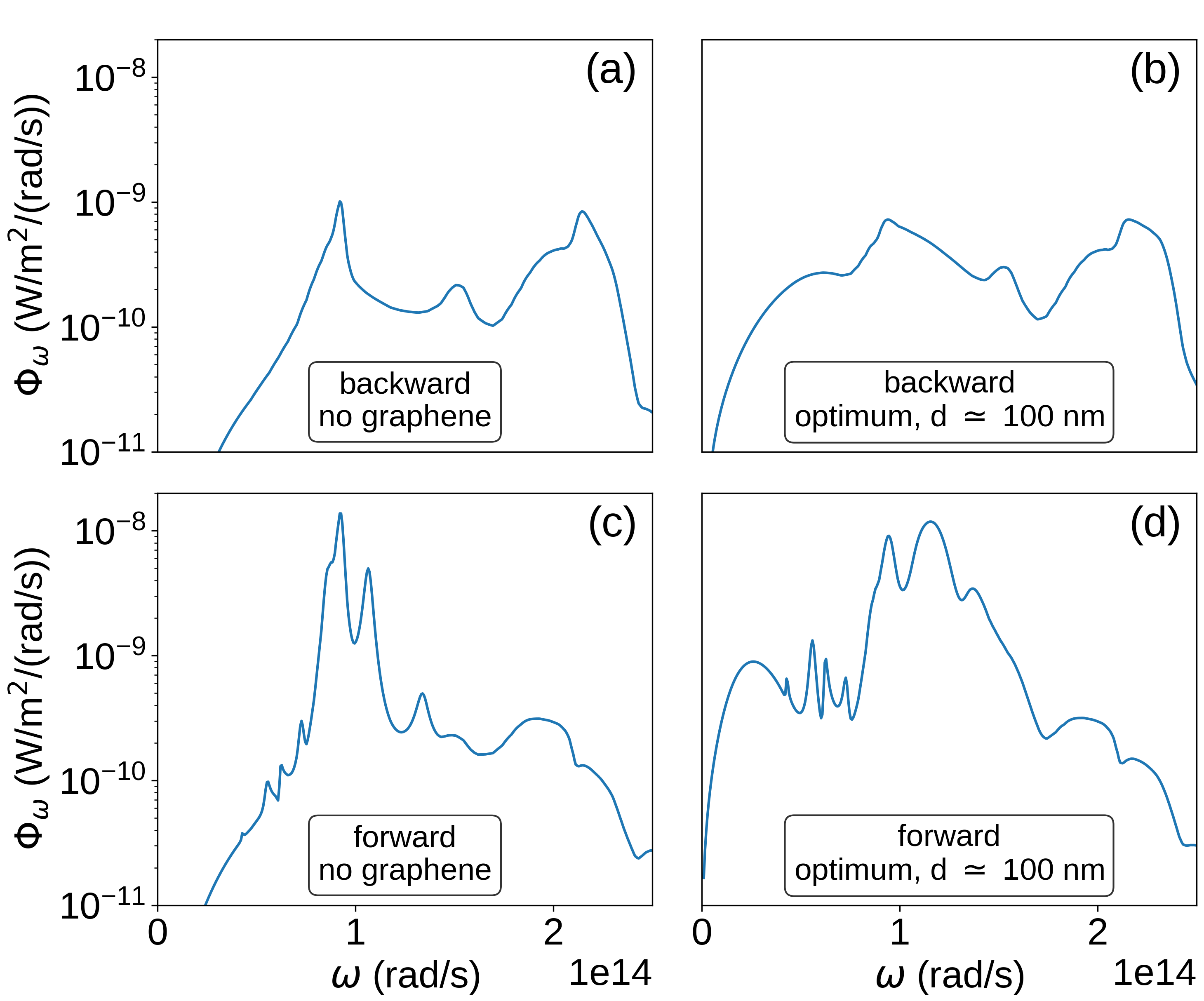}
	\caption{Spectral flux $\Phi_{\omega}$ for the backward temperature gradient [(a), (b)] and for the forward temperature gradient [(c), (d)]. The left column [(a), (c)] correponds to the RTD without graphene, while the right column [(b), (d)] correponds to the RTD with graphene at the optimal chemical potentials for $d \simeq 100$\,nm.}\label{fig:spectre}
\end{figure}

These different effects on $\mathcal{T}_{\text{TM}} (\omega, \mathbf{k})$ have a noticeable impact on the spectral features of the RHT, shown in Fig.~\ref{fig:spectre}. In the first row, the mild effect of graphene on the RHT can be confirmed, as its adjunction only modifies slightly the shape of the two peaks $\omega_{\text{SiO}_2,1}$ and $\omega_{\text{SiO}_2,2}$ of SiO$_2$ and adds a small contribution for low $\omega$. Consequently, the effect of the graphene on $\Phi_{\text B}$ is not substantial. Unlike $\Phi_{\text{B}}$, the addition of graphene at the optimal chemical potentials has a considerable effect on $\Phi_{\text{F}}$. As the broadening in $\omega$ of the resonant modes observed with $\mathcal{T}_{\text{TM}} (\omega, \mathbf{k})$ in Fig.~\ref{fig:tplane} suggested, the spectral flux in the forward scenario is significantly higher on a wide range of $\omega$ (namely from $0.8 \times 10^{14}$\,rad/s to $1.7 \times 10^{14}$\,rad/s). This broadening of the RHT spectrum leads to a significative enhancement of $\Phi_{\text{F}}$ compared to the RTD without graphene. This explains why the adjunction of graphene has such an effect on $\eta$.

\begin{figure}
	\centering
	\includegraphics[width=0.48\textwidth]{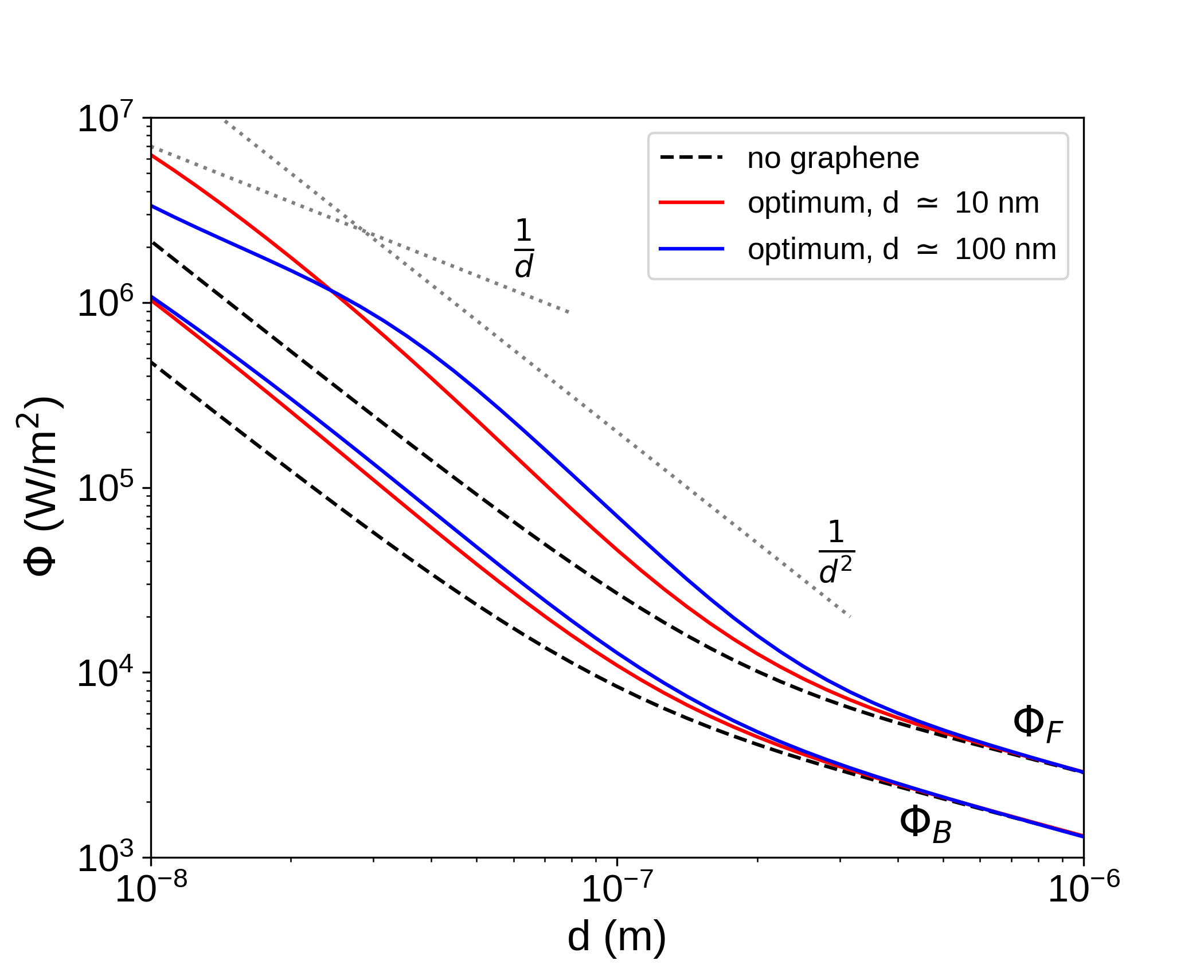}
	\caption{Evolution of the heat fluxes $\Phi_{\text{B}}$ (three lower curves) and $\Phi_{\text{F}}$ (three upper curves) for different configurations. The black dashed lines correspond to the RTD without graphene, the red solid lines to the optimal configuration for $d \simeq 10$\,nm and the blue solid lines to the optimal configuration for $d \simeq 100$\,nm. The two gray dotted lines are visual references for evolutions proportional to $1/d$ and $1/d^2$ as indicated in the plot.}\label{fig:flux}
\end{figure}

It is now interesting to investigate the origin of the non-monotonic behavior of the efficiency increase observed in Fig.~\ref{fig:eta}. As a matter of fact, we notice that for $d < 30$\,nm the optimal $\eta$ decreases when $d$ gets smaller. An explanation to this phenomenon can be found by looking at Fig.~\ref{fig:flux}, where the evolution of $\Phi_{\text{F}}$ and $\Phi_{\text{B}}$ with respect to $d$ is represented for three different configurations: in the absence of graphene (dashed black lines), in the optimal configuration corresponding to $d = 10$\,nm (red curves), and in the optimal configuration corresponding to $d = 100$\,nm (blue curves). We stress that the optimal parameters for the two distances are different, and more specifically we have $\mu_{\text{VO}_2} = 0.16\,$eV and $\mu_{\text{SiO}_2} = 0.09\,$eV for $d = 10$\,nm, whereas $\mu_{\text{VO}_2} = 0.4\,$eV and $\mu_{\text{SiO}_2} = 0.2\,$eV for $d = 100$\,nm.

We start by observing that in the absence of graphene both $\Phi_{\text{F}}$ and $\Phi_{\text{B}}$ present the well-known $1/d^2$ diverging behavior~\cite{Joulain05} of RHT in the near field. On the contrary, in the optimal configuration for $d = 100$\,nm (blue curves), there is a clear transition below 35\,nm towards a $1/d$ behavior, due to the behavior of graphene in the electrostatic regime~\cite{Rodriguez15}. At the same small distances, it is possible to guess the beginning of the same transition in $\Phi_{\text{B}}$, which nevertheless still behaves approximately as $1/d^2$ in the entire range of distances. As a consequence of this transition in power-law behavior, we have an intermediate distance range, containing $d=100\,$nm, where the curve corresponding the forward flux $\Phi_{\text{F}}$ gets farther from the one associated with $\Phi_{\text{B}}$, resulting in the observed increased efficiency. This analysis shows that this transition in the power-law behavior plays indeed a key role in the optimization of the RTD effect. If we keep focusing on the optimal configuration for $d=100\,$nm, we also observe that for smaller distances the transition to a $1/d$ behavior of $\Phi_{\text{F}}$ brings the two fluxes closer to each other, producing then a reduction of efficiency.

If we know switch to the optimal configuration corresponding to $d=10\,$nm, we clearly observe that a deviation from a $1/d^2$ behavior of both fluxes is barely visible. We deduce that the characteristic distance at which this power-law transition takes place depends, not surprisingly, on the values of the two chemical potentials, and more specifically this distance decreases with respect to both chemical potentials. As a consequence, the reduction of the optimal values of $\mu_{\text{SiO}_2}$ and $\mu_{\text{VO}_2}$ for $d=10\,$nm, compared to $d=100\,$nm, is expected in the sense that a transition at smaller distances would be beneficial for the efficiency when reducing $d$. Nevertheless, one should bear in mind that reducing these values is also producing a detuning between the surface modes of the two materials. While this compromise works at intermediate distances such as $d=100\,$nm, it becomes less efficient when reducing the separation distance, explaining the non-monotonic behavior of the optimized $\eta$ in Fig.~\ref{fig:eta} as a function of the distance.

In conclusion, we have highlighted an enhancement of the efficiency of the radiative thermal rectification in near field using tunable graphene sheets to mediate the heat transfer between a polar  and a phase-change material. We have shown that this enhancement results from a better coupling between the surface modes supported by the the two interacting solids thanks to the presence of graphene sheets, and connected this effect to a transition in the power-law dependence of heat flux from a $1/d^2$ to a $1/d$ behavior. Our findings shed further light on the remarkable potential of graphene in the domain of active thermal management.

\section*{Author declarations}

\subsection*{Conflict of interest}

The authors have no conflicts to disclose.

\section*{Data availability}

The data that support the findings of this study are available from the corresponding authors upon reasonable request.

\end{document}